\documentclass[letterpaper]{article}
\usepackage{iccc}

\usepackage{times}
\usepackage{helvet}
\usepackage{courier}
\usepackage{multirow}
\usepackage{amssymb}
\usepackage{amsmath}
\usepackage{pifont}
\usepackage{microtype}
\usepackage{url}
\usepackage{booktabs}
\usepackage{amsfonts}
\usepackage{xcolor}         % colors
\usepackage{graphicx}
\usepackage{url}
\usepackage{subcaption}  
\usepackage{svg}

\usepackage{xspace,mfirstuc,tabulary}
\usepackage{listings}

\definecolor{codegreen}{rgb}{0,0.6,0}
\definecolor{codegray}{rgb}{0.5,0.5,0.5}
\definecolor{codepink}{RGB}{252, 142, 172}
\definecolor{codepurple}{rgb}{0.58,0,0.82}
\definecolor{backcolour}{RGB}{245,245,245}
\lstdefinestyle{mystyle}{
    backgroundcolor=\color{backcolour},   
    commentstyle=\color{magenta},
    keywordstyle=\color{blue},
    numberstyle=\tiny\color{codegray},
    stringstyle=\color{codepurple},
    basicstyle=\fontfamily{\ttdefault}\footnotesize,
    breakatwhitespace=false,         
    breaklines=true,                 
    keepspaces=true,    
    frame=single,
    numbersep=5pt,                  
    showspaces=false,                
    showstringspaces=false,
    showtabs=false,                  
    tabsize=2,
    classoffset=1, % starting new class
    keywordstyle=\color{violet},
    classoffset=0,
}
\lstdefinelanguage{JavaScript}{
  keywords={typeof, new, true, false, catch, function, return, null, catch, switch, var, if, in, while, do, else, case, break},
  keywordstyle=\color{blue}\bfseries,
  ndkeywords={class, export, boolean, throw, implements, import, this},
  ndkeywordstyle=\color{darkgray}\bfseries,
  identifierstyle=\color{black},
  sensitive=false,
  comment=[l]{//},
  morecomment=[s]{/*}{*/},
  commentstyle=\color{purple}\ttfamily,
  stringstyle=\color{red}\ttfamily,
  morestring=[b]',
  morestring=[b]''
}

\title{Reimagining Dance: Real-time Music Co-creation between Dancers and AI}
 \author{
   Olga Vechtomova$^*$ 
   \and
   Jeff Bos$^{**}$
   \\
   \ \\
   $^*$University of Waterloo, Canada \\
   $^{**}$WordSynth Inc. \\
   \texttt{ovechtom@uwaterloo.ca}, \texttt{jeff@wordsynth.com}
   \\
 }

\begin{document} 
\maketitle
\begin{abstract}
Dance performance traditionally follows a unidirectional relationship where movement responds to music. While AI has advanced in various creative domains, its application in dance has primarily focused on generating choreography from musical input. We present a system that enables dancers to dynamically shape musical environments through their movements. Our multi-modal architecture creates a coherent musical composition by intelligently combining pre-recorded musical clips in response to dance movements, establishing a bidirectional creative partnership where dancers function as both performers and composers. Through correlation analysis of performance data, we demonstrate emergent communication patterns between movement qualities and audio features. This approach reconceptualizes the role of AI in performing arts—as a responsive collaborator that expands possibilities for both professional dance performance and improvisational artistic expression across broader populations.
\end{abstract}

\section{Introduction}
Dance and music traditionally exist in a hierarchical relationship where movement follows sound. Typically, choreographers design dance to existing music, or collaborate with composers to create accompanying scores. Even in improvisational dance, performers respond to pre-composed or live music, but rarely influence the musical composition itself.

Artificial intelligence now offers an opportunity to invert this relationship. While most AI systems in dance maintain the traditional paradigm by generating choreography from musical input, our research proposes a fundamental shift: enabling dancers to dynamically shape musical environments through their movements. This approach reconceptualizes dancers as both performers and composers, establishing a bidirectional creative partnership between human movement and AI-created sound.

\begin{figure}[htbp]              % h: here, t: top, b: bottom, p: page of floats
  \centering                      % center the figure
  \includegraphics[width=0.45\textwidth]{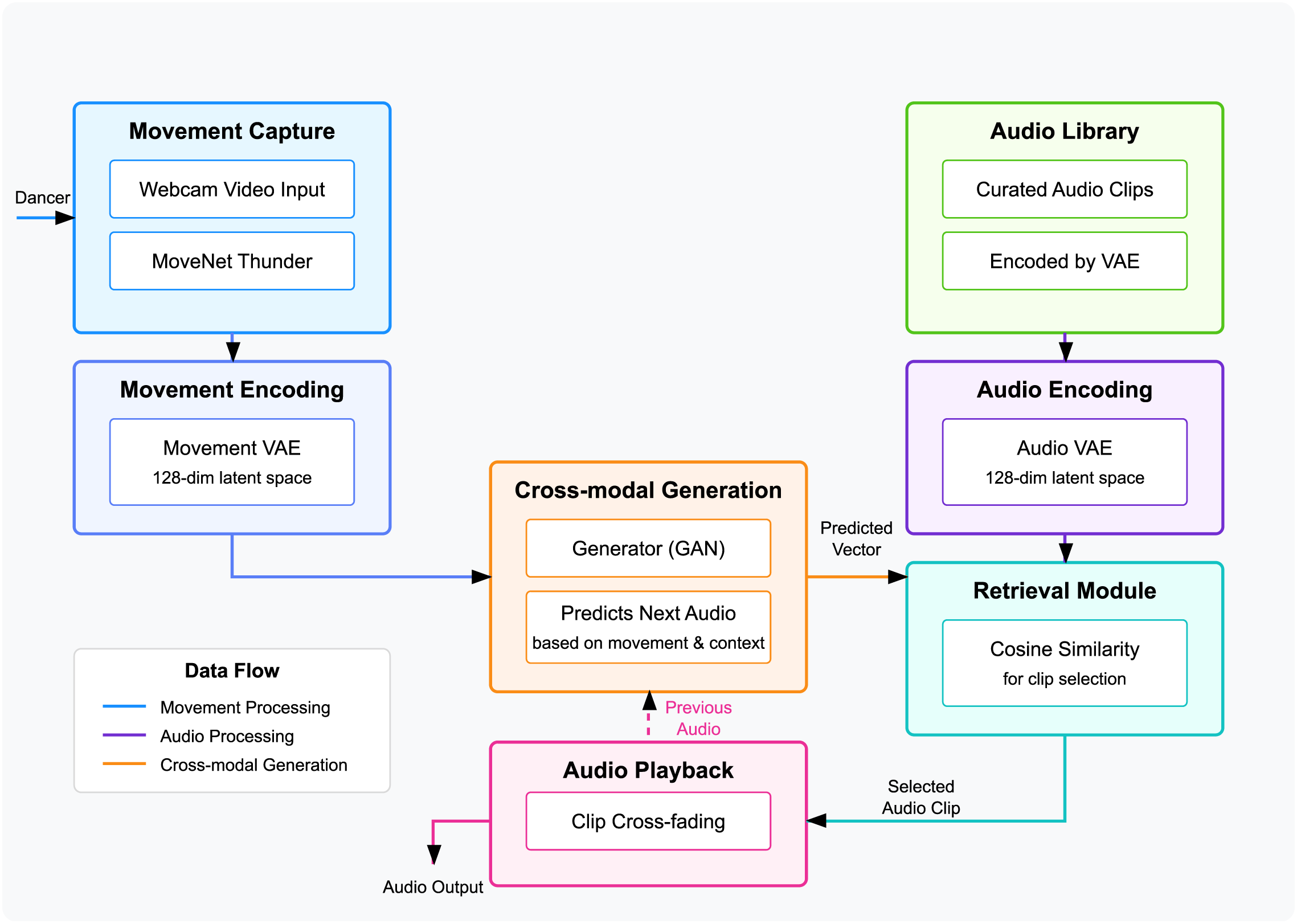}
  \caption{System diagram}
  \label{fig:system-diagram}
\end{figure}

In this paper, we present a system that enables dancers to dynamically influence musical composition in real-time through their movements. Our technical contribution is a multi-modal architecture (Figure \ref{fig:system-diagram}) that selects and seamlessly combines pre-recorded musical clips in response to a dancer's movement patterns. Unlike previous approaches, our system creates an evolving musical environment where the dancer becomes an active co-creator of the soundscape rather than merely responding to it. We demonstrate through correlation analysis of  pilot performances that this approach creates emergent communication patterns between dancer and system, establishing meaningful bidirectional relationships between specific movement qualities and audio features.

This approach not only transforms dance performance but also opens new theoretical and practical directions for studying creative partnership between human performers and AI systems. While considerable research has examined AI's role in generating static artistic content like images or poetry, the dynamic, real-time nature of dance performance presents unique challenges that remain largely unexplored. %The gap in our current understanding is particularly acute in live performance contexts, where the immediacy of response and quality of interaction are crucial to artistic expression. While considerable research has examined AI's role in generating static artistic content like images or composed music, the dynamic, real-time nature of dance performance presents unique challenges that remain largely unexplored.

\section{Related Work}
\label{sec:related}
Existing research in dance and AI has primarily focused on generating choreographic movements from musical input. Tang et al. \shortcite{tang2018dance} developed an LSTM-autoencoder model that synthesizes dance choreography by mapping acoustic features to motion features, addressing the challenge of selecting appropriate dance figures that match musical elements. Lee et al. \shortcite{lee2019dancing} proposed a synthesis-by-analysis framework that decomposes dance into basic units to generate style-consistent and beat-matching movements from music. While these approaches demonstrate technical sophistication in movement generation, they maintain the traditional unidirectional relationship where dance follows music.
%This approach is consistent with broader trends in AI creativity research, where cross-modal generation systems typically follow established creative hierarchies rather than challenging them. Recent work in music-to-dance or text-to-image generation has shown impressive technical capabilities, but often reinforces existing creative processes rather than reimagining the relationships between different modalities and creative agents.

Some researchers have begun exploring more interactive approaches to AI in dance. Kumar et al. \shortcite{kumar2020creativity} developed LuminAI, an improvisational dance installation where an AI agent dances with users, implementing a lead-and-follow dynamic based on creativity metrics. The choreographic duo A$\Phi$E's recent ``Lilith.Aeon'' performance, as reported in The Guardian \cite{winship2024small}, demonstrated how an AI system trained on human-generated movements could become an active creative partner, suggesting new movement possibilities while maintaining the distinctive style of the choreographers. The Royal Ballet choreographer Wayne McGregor's collaboration with Google Arts \& Culture Lab produced AISOMA, a system that suggests new movement variations by analyzing rehearsal videos, expanding the choreographic possibilities available to dancers and choreographers \cite{winship2024small}.

%However, these systems still primarily operate in a choreographic rather than performance context, focusing on movement generation rather than real-time musical interaction. The few systems that do engage with real-time performance typically maintain the traditional unidirectional relationship where AI analyzes music to suggest or generate dance movements. This approach, while valuable, fails to exploit the full potential of AI as a collaborative partner in live performance, particularly in terms of allowing dancers to shape the musical environment through their movements.

\textbf{Theoretical Foundations.} This research builds on several theoretical foundations that help frame our understanding of human-AI creative collaboration. Particularly relevant is the concept of ``mixed-initiative creative interfaces'' developed by Deterding et al. \shortcite{deterding2017mixed}, which describes systems where human and computational agents take turns contributing to an evolving artistic work. In the context of dance performance, this framework takes on new dimensions as the collaboration happens in real-time, with the dancer's physical movements and the AI's musical responses creating a dynamic feedback loop.
Additionally, our work is informed by computational creativity concepts such as Colton and Wiggins' \shortcite{colton2012computational} "creative responsibility," where AI systems take on creative roles beyond mere tools—evaluating aesthetics and inventing processes. This complements Jennings' \shortcite{jennings2010developing} notion of ``creative autonomy,'' which requires systems to independently evaluate and evolve their standards. 

While the pilot study reported in our paper establishes technical foundations, these frameworks guide our vision for AI systems that can function as genuine creative partners in dance performance.

Beyond computational creativity, performance studies literature on improvisation and real-time creative decision-making provides another crucial theoretical perspective. Foundational work by Bailey \shortcite{bailey1992improvisation} and Nachmanovitch \shortcite{nachmanovitch1990free} established key principles of improvisational practice, while recent scholarship addresses the complexities of dance improvisation and human-machine interaction. De Spain's \shortcite{despain2014landscape} topographical approach illuminates how dancers make moment-to-moment decisions, and Foster \shortcite{foster2002dances} examines how improvisational structures emerge through real-time choreographic choices. For human-AI creative collaboration specifically, Hoffman and Weinberg's \shortcite{hoffman2011interactive} work on interactive robotic improvisation offers frameworks for understanding how performers adapt to non-human partners. Carter's \shortcite{carter2000improvisation} analysis of improvisation as breaking established conventions to discover new artistic expressions that "could not be found in a systematic preconceived process" is particularly relevant. Following Carter, our system aims to create new paradigms that enable real-time invention and discovery through the act of creation itself.

%Performance studies literature, particularly work on improvisation and real-time creative decision-making, provides another crucial theoretical perspective. While foundational work by Bailey \cite{bailey1992improvisation} and Nachmanovitch \cite{nachmanovitch1990free} established key principles of improvisational practice, more recent scholarship has specifically addressed the complexities of dance improvisation and human-machine interaction. De Spain's \cite{despain2014landscape} topographical approach to movement improvisation provides insights into how dancers make moment-to-moment decisions, while Foster \cite{foster2002dances} examines how improvisational structures emerge through real-time choreographic choices. In the context of human-AI creative collaboration, Hoffman and Weinberg's \cite{hoffman2011interactive} work on interactive robotic improvisation offers valuable frameworks for understanding how performers adapt to non-human partners. Equally crucial is Carter's \cite{carter2000improvisation} analysis of improvisation as a means of breaking established conventions and discovering genuinely new artistic forms of expression that ``could not be found in a systematic preconceived process.'' Following Carter, our AI system aims to create new paradigms, allowing for real-time invention and discovery through the act of creation itself.
    
\section{System Architecture}
Our system enables real-time generation of responsive musical accompaniment to dance movements through a multi-stage machine learning pipeline. The architecture comprises three primary components: (1) an audio encoding/decoding system, (2) a movement encoding system, and (3) a cross-modal generation network that predicts appropriate musical responses to movement. These components work in concert to create a cohesive interactive performance environment.

\textbf{Audio Representation Learning.}
To learn audio representations, we trained a Variational Autoencoder (VAE) \cite{kingma2014auto} on spectrograms of 3.5-second audio clips. The audio VAE consists of convolutional layers for both encoding and decoding, with a 128-dimensional latent space representation. This architecture effectively compresses spectrograms into a compact latent code that preserves meaningful acoustic properties while discarding noise. The encoder uses five convolutional layers with ReLU activations to transform input spectrograms (224×224×1) into a latent distribution, while the decoder reverses this process through transposed convolutions.

\begin{figure}[htbp]
  \centering
  % first subfigure
  \begin{subfigure}[b]{0.28\textwidth}
    \centering
    \includegraphics[width=\linewidth]{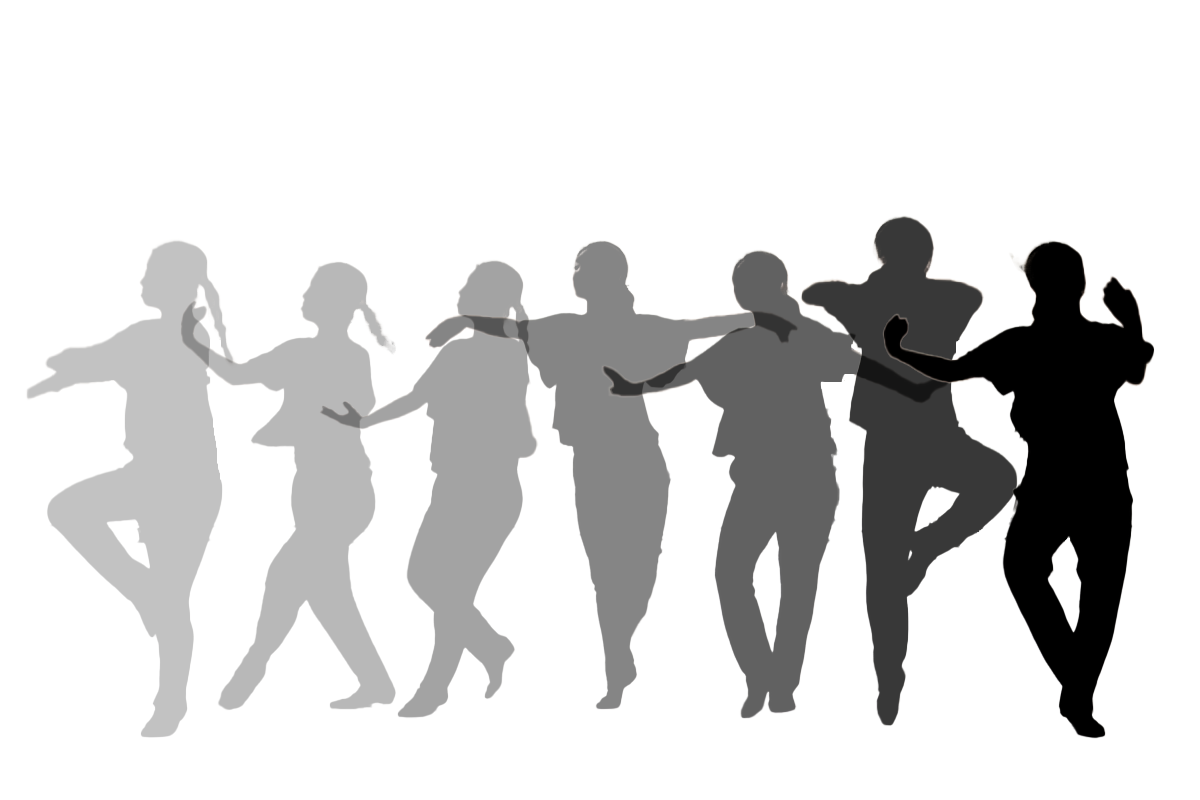}
    \caption{Dance timelapse photo}
    \label{fig:dance-timelapse}
  \end{subfigure}
  \hfill
  % second subfigure, with bottom cropped by 2cm
  \begin{subfigure}[b]{0.18\textwidth}
    \centering
    \includegraphics[width=\linewidth,
                     trim=0 0.8cm 0 0,clip]{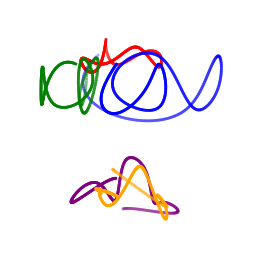}
    \caption{Movement trajectories}
    \label{fig:movement-trajectories}
  \end{subfigure}

  \caption{(a) Timelapse photo of dancer's movements. (b) Trajectories of dancer's movements captured by our system. The landmarks are colour-coded as follows: red - head, green - left wrist, blue - right wrist, orange - left ankle, magenta - right ankle.}
  \label{fig:example-trajectories}
\end{figure}

\textbf{Movement Representation Learning.}
To encode dance movements, we developed a parallel VAE architecture that processes visual representations of movement trajectories. The movement data is collected by using TensorFlow MoveNet Thunder pipeline, which analyzes either pre-recorded videos during training or webcam video stream during inference. Rather than working with raw skeletal joint data, we first transform movement sequences into color-coded trajectory images (Figure \ref{fig:movement-trajectories}). Each 3.5-second movement sequence is represented as an RGB image (256×256×3) where five key landmarks (head, left wrist, right wrist, left ankle, right ankle) are visualized as coloured curves showing their trajectories over time.

This image-based approach allows us to leverage convolutional architectures commonly used in computer vision while capturing the temporal dynamics of movement. The movement VAE employs a structure parallel to the audio VAE, with the encoder producing a 128-dimensional latent vector that encapsulates the essential spatial and temporal characteristics of the dance movement.

\textbf{Cross-Modal Generation}
The Generative Adversarial Network (GAN) \cite{goodfellow2014generative} bridges the movement and audio domains. The GAN's generator takes two inputs: (1) the latent representation of the current movement and (2) the latent representation of the previous audio clip. It then predicts the latent vector for the next audio clip that would best complement the current movement.

The generator employs a latent combiner module that integrates movement and audio latent vectors. While we experimented with several combination methods (concatenation, multiplication, and various learned approaches including gated, FiLM, and cross-attention mechanisms), we found that pointwise addition produces the most effective results. This addition operation is followed by a multi-layer network with hidden dimensions of 256 units, LayerNorm for stabilization, and LeakyReLU activations.

\textbf{Retrieval Module}
Rather than directly decoding the predicted latent vector, which could result in lower audio quality, we employ a retrieval-based approach. We calculate the cosine similarity between the predicted latent vector and the latent representations of clips in our reference database. The audio clip with the highest similarity is selected, ensuring high-quality output while maintaining contextual relevance.

\begin{figure}[htbp]
  \centering
  % Row 1: (a) and (b)
  \begin{subfigure}[b]{0.48\columnwidth}
    \centering
    \fcolorbox{black}{white}{%
      \includegraphics[width=\linewidth]{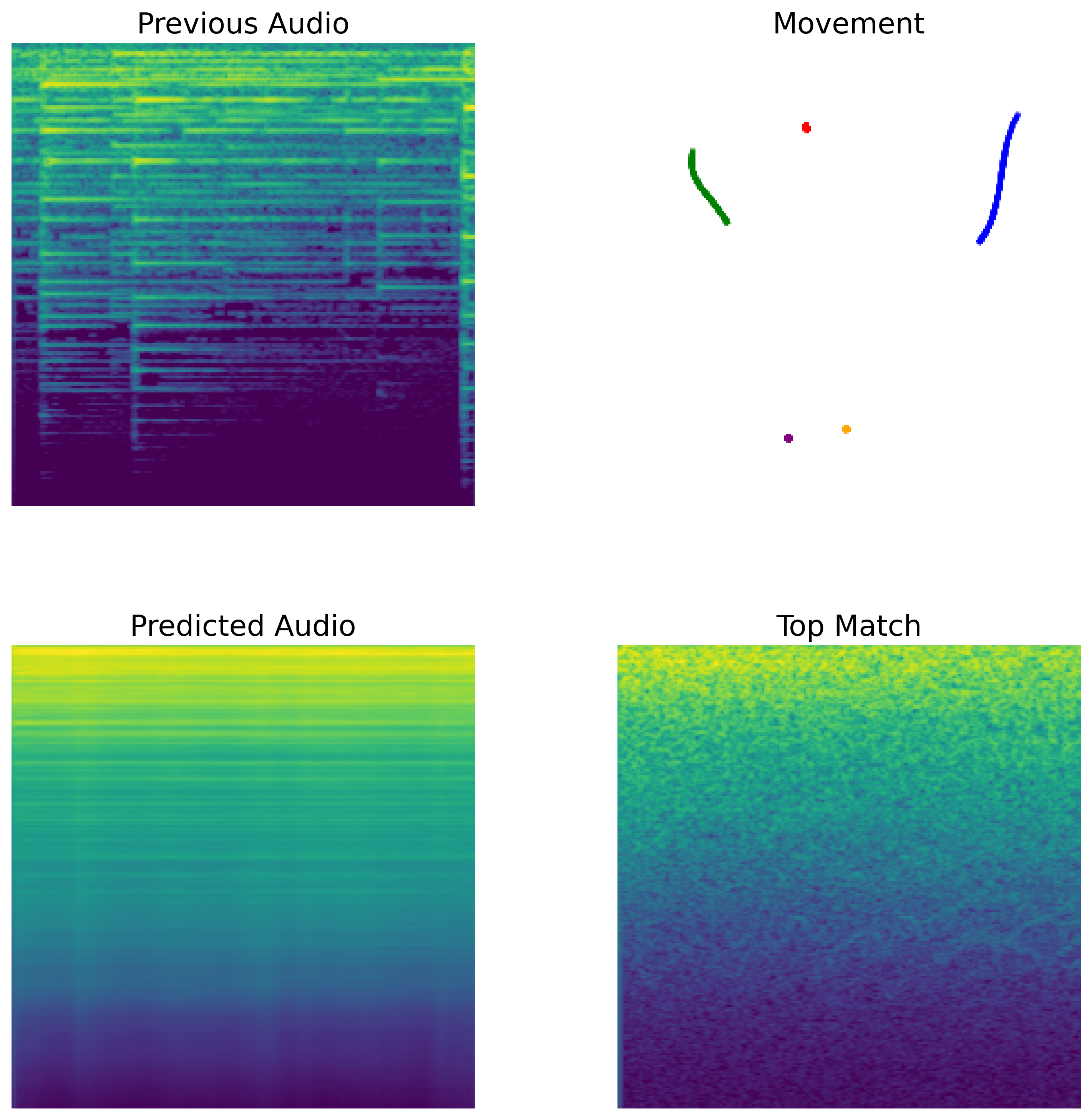}%
    }
    \caption{}
    \label{fig:modalities-effect-a}
  \end{subfigure}
  \hfill
  \begin{subfigure}[b]{0.48\columnwidth}
    \centering
    \fcolorbox{black}{white}{%
      \includegraphics[width=\linewidth]{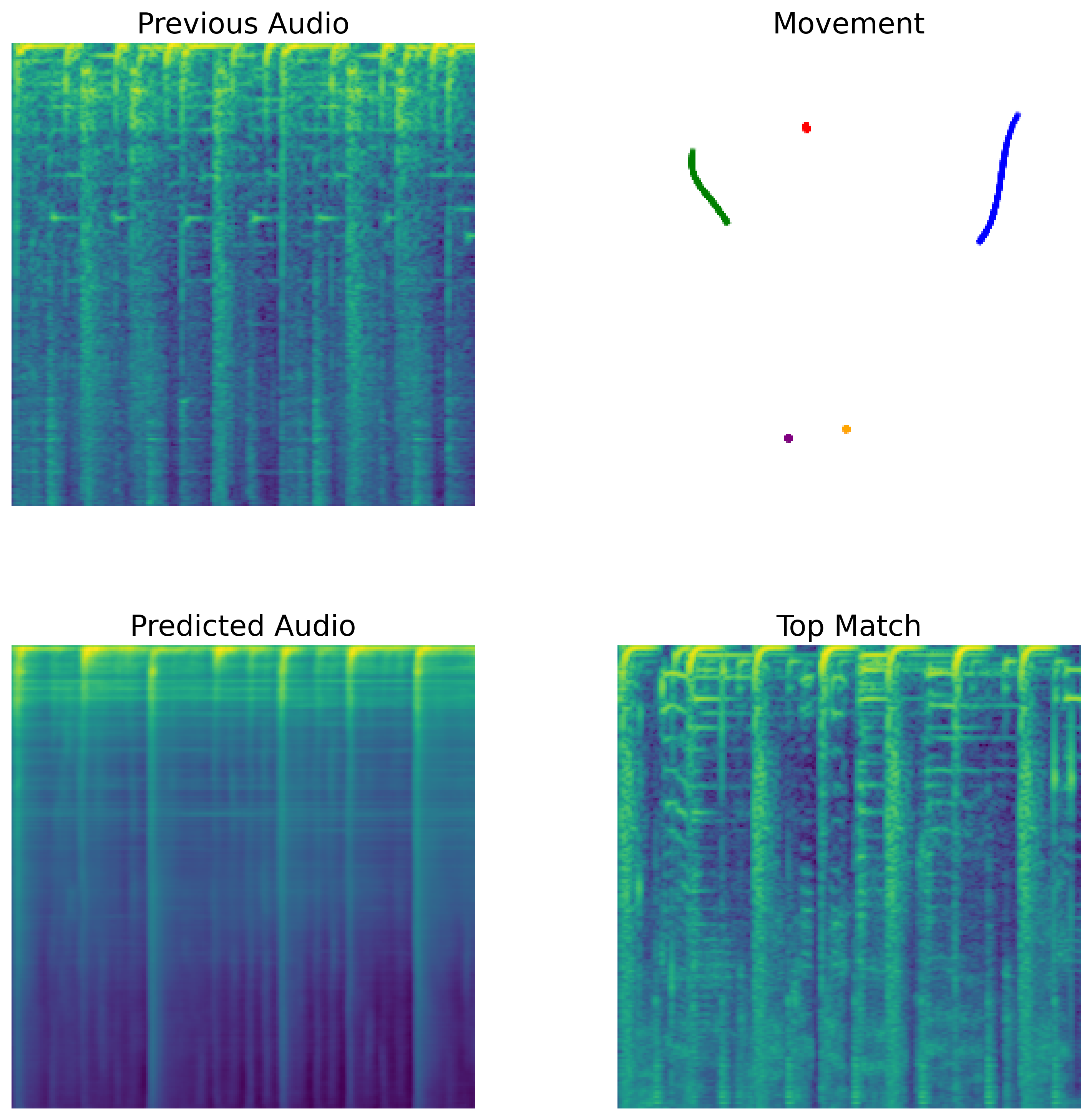}%
    }
    \caption{}
    \label{fig:modalities-effect-b}
  \end{subfigure}

  \vspace{1em} % vertical separation between rows

  % Row 2: (c) and (d)
  \begin{subfigure}[b]{0.48\columnwidth}
    \centering
    \fcolorbox{black}{white}{%
      \includegraphics[width=\linewidth]{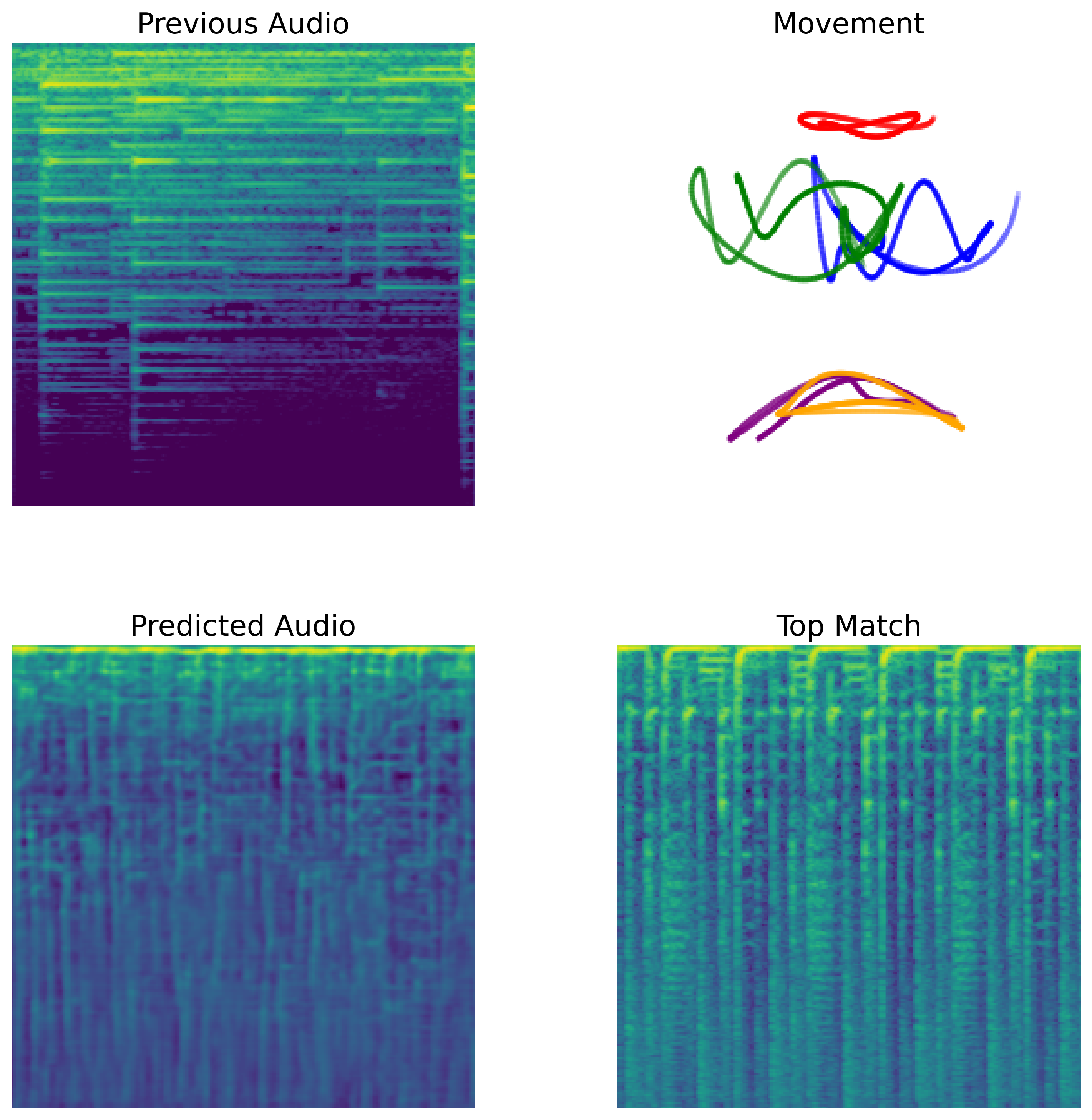}%
    }
    \caption{}
    \label{fig:modalities-effect-c}
  \end{subfigure}
  \hfill
  \begin{subfigure}[b]{0.48\columnwidth}
    \centering
    \fcolorbox{black}{white}{%
      \includegraphics[width=\linewidth]{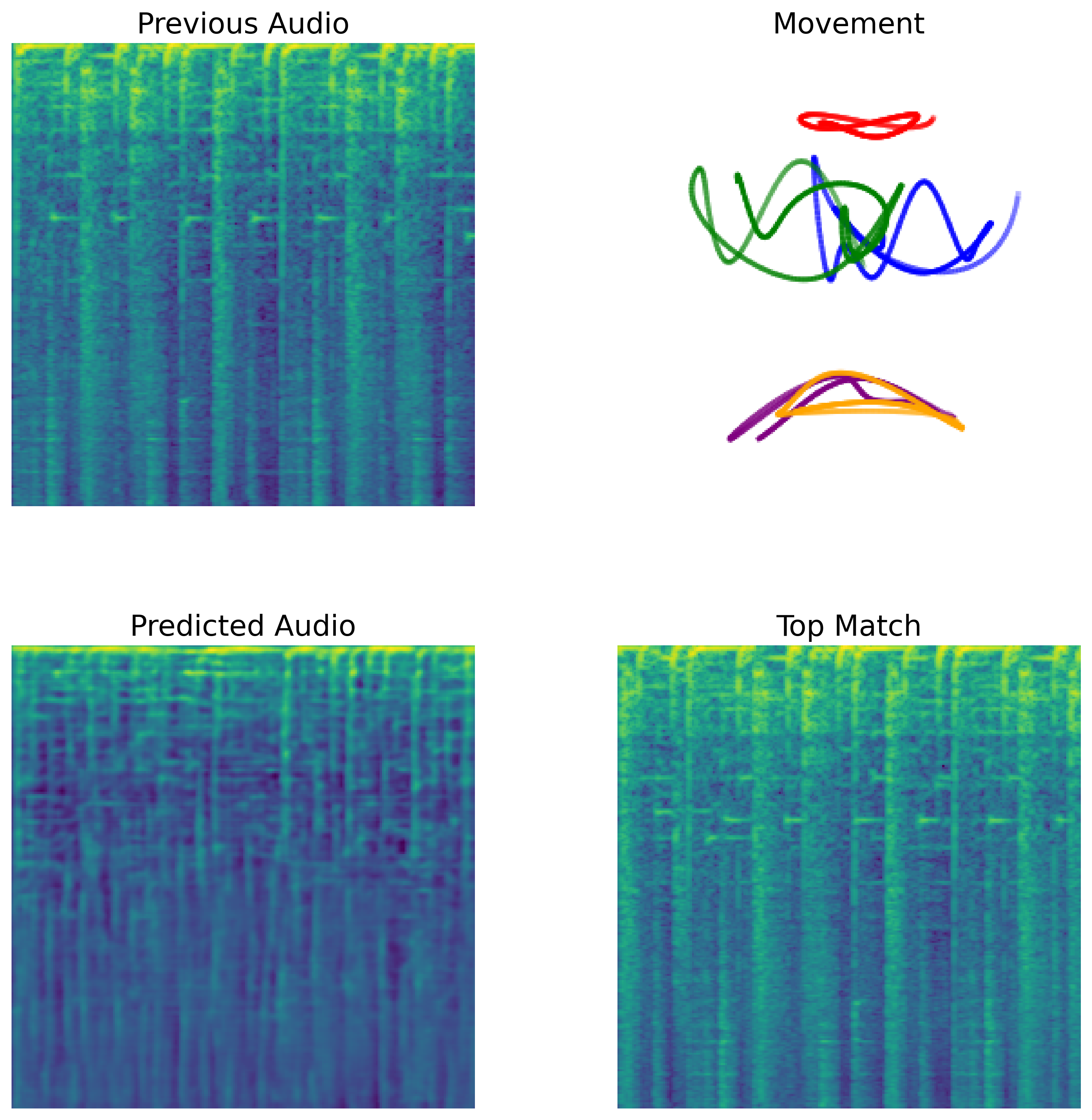}%
    }
    \caption{}
    \label{fig:modalities-effect-d}
  \end{subfigure}

  \caption{System response examples showing the influence of both inputs on prediction.} %Each panel displays: previous audio (top left), movement data (top right), predicted audio (bottom left), and selected audio (bottom right). Panels demonstrate: (a) minimal movement generating ambient sounds; (b) same movement with different previous audio yielding new predictions; (c) energetic break-dancing producing rhythmic selections; (d) same energetic movement with different previous audio producing varied rhythmic outcomes.}
  \label{fig:modalities-effect}
\end{figure}

Figure \ref{fig:modalities-effect} shows how both previous audio and movement influence the system's predictions. With identical previous audio but different movements, the system selects dramatically different clips: ambient music for minimal movement (\ref{fig:modalities-effect-a}) versus rhythmic clips for energetic break-dancing (\ref{fig:modalities-effect-c}). Similarly, when movement remains constant but previous audio changes (\ref{fig:modalities-effect-a} vs. \ref{fig:modalities-effect-b}), the predicted clips also differ.

%Figure \ref{fig:modalities-effect} demonstrates how the system's prediction and subsequent selection of the next clip is influenced by both inputs: previous audio clip and the movement. For example, comparing \ref{fig:modalities-effect-a} and \ref{fig:modalities-effect-c}, we see that with the same previous audio clip, but different movements, the system recommends dramatically different outcomes: sustained ambient clip when the dancer is almost still (\ref{fig:modalities-effect-a}) and rhythmic clip when the dancer is performing a break-dance routine (\ref{fig:modalities-effect-c}). When we fix the conditioning movement input, but change the previous clip (e.g., \ref{fig:modalities-effect-a} vs. \ref{fig:modalities-effect-b}), we also see different predicted clips.

\textbf{Real-time Inference}
We developed a React/NodeJS application, which operates as follows during live performance:

\textit{Movement Capture:} A webcam captures the dancer's movements, which are processed by a pose estimation model (TensorFlow MoveNet Thunder) to extract the five key landmarks.

\textit{Movement Encoding:} The landmarks' trajectories are rendered as a color-coded image and encoded by the movement VAE into a latent vector.

\textit{Audio Prediction:} The movement latent vector and the previous audio clip's latent vector are fed into the GAN, which predicts the next audio clip's latent representation.

\textit{Clip Selection:} The system retrieves the audio clip whose latent representation has the highest cosine similarity to the predicted vector.

\textit{Audio Playback:} The selected clip is cross-faded with the currently playing audio to create seamless transitions.

The dancer can also curate the source audio library from which the system selects clips, allowing performers to influence the overall sonic palette and musical style of their movement-conditioned compositions.

%This pipeline executes with low latency, creating a responsive system where the music follows the dancer's movements while maintaining musical coherence. The result is a dynamic performance environment where the dancer effectively composes the musical score through movement, inverting the traditional relationship between dance and music.

\textbf{Dataset.} The aligned video-audio dataset used to train the movement VAE and GAN consists of 18K 3.5-second recordings, 17K of which were sourced from the AIST dance dataset \cite{aist-dance-db} and 1K from our own dataset containing video material recorded specifically for this project or provided to us by professional dance collaborators. The audio VAE was trained on a larger set of 50K audio clips from the authors' studio recordings, spanning multiple genres of electro-acoustic music with varying tempos and instrumentation. %This diverse dance-style dataset ensures the system can respond to a wide range of movement vocabularies during performance.

\section{Pilot study}
\label{sec:Analysis}

We conducted a pilot study with three participants of varying dance experience: P1 (10+ years ballet), P2 (2-3 years ballet/jazz), and P3 (no formal training). Each dancer performed improvisational movement with the system for up to 30 minutes. We recorded both video and system-generated audio, collecting over 70 minutes of data. Our analysis aimed to identify relationships between dance movements and generated audio, particularly examining which audio features correlate most strongly with movement intensity.

\textbf{Video and Audio Features.}
We segmented performances into 10-second clips and extracted movement data using MoveNet Thunder to track key body points (head, wrists, ankles) normalized to [-1, 1]. Movement energy was measured as the Euclidean distance between corresponding points in consecutive frames, with statistical measures (mean, min, max, standard deviation) computed for each clip. For audio, we extracted 47 features using Librosa and Essentia, including spectral features (MFCCs, contrast, flux), chroma features, and psychoacoustic measures. Key audio features in our analysis include \textit{MFCCs} (representing timbre, with \textit{mfcc\_1} capturing overall spectral shape), \textit{spectral contrast} (peak-valley differences across frequency bands), \textit{chroma} (pitch class distribution), and \textit{spectral flux} (frame-to-frame spectral changes).

\textbf{Statistical Analysis\footnote{
  Detailed statistical analysis results are provided on the
  supplementary‐material website:
  https://sites.google.com/view/reimagining-dance
}}
To explore the relationship between dance movement and system-generated audio, we applied several statistical methods. Pearson correlation analysis was performed to assess the linear relationships between individual movement energy measures (average, maximum, minimum, and standard deviation) and audio features. Principal Component Analysis (PCA) was conducted separately on video and audio features to identify patterns of variance and reduce dimensionality, while Canonical Correlation Analysis (CCA) was used to examine the multivariate relationships between movement and audio feature spaces.

Additionally, we used Partial Least Squares (PLS) regression to model predictive relationships between the two modalities, evaluating how well sets of audio features could predict movement energy metrics, and vice versa. To identify the most influential audio features, we employed Random Forest regression models, computing feature importance scores based on their contribution to predicting each movement energy statistic. The quality of these predictive models was evaluated using the coefficient of determination (R²).

Together, these analyses allowed us to identify which audio features were most strongly associated with variations in dancers' movement intensity and to assess the strength of the coupling between movement and audio dynamics produced by the system.

\textbf{Results.} Our analysis revealed several key relationships between movement energy and audio features generated by the system.

\textit{Correlation Analysis.}
Pearson correlation analysis showed that the minimum movement energy (\textit{min\_energy}) had the strongest and most consistent relationships with audio features. In particular, \textit{mfcc\_1} (first Mel-frequency cepstral coefficient) exhibited a significant negative correlation with min\_energy (r = -0.45, p \textless 0.001), suggesting that clips with lower minimum movement energy were associated with audio segments characterized by smoother spectral shapes. Spectral contrast in the sixth frequency band (\textit{spec\_contrast\_6}) and \textit{mfcc\_7} also showed significant positive correlations with min\_energy.

\textit{Principal Component Analysis (PCA).}
PCA indicated that a small number of components captured substantial variance in both movement and audio features. In particular, variations in energy-based movement measures (average, minimum, maximum, and standard deviation) loaded heavily onto the first few principal components, while audio variance was dominated by MFCCs and spectral features.

\textit{Canonical Correlation Analysis (CCA).}
CCA revealed moderate canonical correlations between the combined movement energy statistics and audio feature sets. The first canonical component pair linked high standard deviation of movement energy with variations in spectral complexity and dissonance in the audio, indicating multivariate coupling between movement expressivity and audio texture.

\textit{Partial Least Squares (PLS) Regression.}
PLS regression models found that audio features could modestly predict movement energy, with the highest R² value (0.103) for predicting \textit{min\_energy}. Conversely, movement features showed stronger predictive power for certain audio characteristics: movement energy metrics could predict \textit{mfcc\_1} with an R² of 0.202 and \textit{mfcc\_7} with an R² of 0.162, suggesting that dancers’ movement dynamics influenced the timbral qualities of the generated soundtrack.

\textit{Random Forest Feature Importance.}
Random forest regressions further confirmed the importance of specific audio features. \textit{mfcc\_1}, \textit{mfcc\_7}, \textit{spec\_contrast\_6}, and \textit{spectral\_flux} consistently emerged as the most important predictors of movement energy statistics across models, aligning with the findings of the linear analyses.

Overall, the results suggest that the soundscape created by our system responded most consistently to variations in dancers’ minimum movement energy, with audio features related to timbre and spectral dynamics (MFCCs and spectral contrast) showing the strongest associations with movement intensity.

\textbf{Qualitative results}
Dancers reported a fluid exchange of initiative with the system throughout their performances. The system often influenced their movement choices at both macro and micro levels, inspiring exploration of new dance expressions to discover corresponding musical responses. Conversely, dancers sometimes deliberately attempted to redirect the musical atmosphere, such as introducing energetic movement during ambient passages. While the system typically required 5-10 seconds to adapt to significant changes in dance energy, this delay was intentionally designed to maintain musical coherence. This adaptation period could potentially be customized based on dance genre preferences, balancing responsiveness against musical continuity.

\section{Conclusion}
This research presents a novel approach to dance through AI-mediated co-creation, fundamentally reimagining the traditional relationship between movement and music. Our system enables dancers to dynamically shape musical environments while simultaneously responding to them, creating a bidirectional creative partnership where initiative flows fluidly between human and machine. Statistical analysis confirms meaningful correlations between specific movement qualities and audio features. 

This work explores a multi-layered creative relationship. The original composer's musical intent—which may itself be improvisational—becomes raw material for a new emergent composition, dynamically rearranged and remixed through the dancer's movements. The resulting sonic experience is a form of real-time collage where three creative forces converge: the original compositional elements, the system's algorithmic decision-making, and the dancer's embodied expression. Each performance thus represents a unique relationship between these creative entities, with the dancer physically sculpting a new musical composition from fragments of the composer's work, creating something neither could have produced independently.

By inverting the traditional paradigm where movement follows sound, we position dancers as active co-creators of the musical arrangements. The dance movements can serve as a novel compositional tool for creating musical content that has artistic value beyond the performance context. This approach offers new creative possibilities not only to dance performance, but also as a form of musical composition.

While the system was primarily envisioned as a co-creative artistic framework for dance performance, another promising application emerged during our pilot study. We observed that when participant dancers experienced fatigue and naturally reduced their movement intensity, the system organically transitioned from dynamic rhythmic music to more ambient soundscapes. This adaptive quality could be valuable not only in dance performance, but also in exercise and training settings.

We are currently planning a large-scale study with a professional dance company to evaluate the system's impact on choreographic process and audience reception. Through this ongoing research, we aim to develop a deeper understanding of the emergent creative language that evolves between human dancers and AI musical collaborators.
%This research presents a new approach to dance through AI-mediated co-creation. Our system enables dancers to shape musical environments while simultaneously responding to them, creating a bidirectional partnership where initiative flows between human and machine. Statistical analysis confirms correlations between movement qualities and audio features, while dancer feedback highlights this fluid exchange during performances. We are currently planning a large-scale study with a professional dance company.
%Our findings suggest emergent communication patterns in our dance-music co-creation system. The strong correlations with timbral features suggest these sound qualities served as primary ``communication channels'' between the dancer in our study and the system. 
%The complex correlations observed represent ``attractors'' in the co-creative space - specific combinations of movement and audio features that the dancer and system naturally converged toward during interaction. This demonstrates that the dancer and the system establish a bidirectional creative relationship where the dancer shapes the musical environment through movement while simultaneously responding to the evolving soundscape. 

\bibliographystyle{iccc}
\bibliography{iccc}

\begin{thebibliography}{}

\bibitem[\protect\citeauthoryear{Bailey}{1992}]{bailey1992improvisation}
Bailey, D.
\newblock 1992.
\newblock {\em Improvisation: Its Nature and Practice in Music}.
\newblock Da Capo Press.

\bibitem[\protect\citeauthoryear{Carter}{2000}]{carter2000improvisation}
Carter, C.~L.
\newblock 2000.
\newblock Improvisation in dance.
\newblock {\em Journal of Aesthetics and Art Criticism} 58(2):181--190.

\bibitem[\protect\citeauthoryear{Colton and Wiggins}{2012}]{colton2012computational}
Colton, S., and Wiggins, G.~A.
\newblock 2012.
\newblock Computational creativity: The final frontier?
\newblock In {\em ECAI 2012}, volume 242 of {\em Frontiers in Artificial Intelligence and Applications},  21--26.
\newblock IOS Press.

\bibitem[\protect\citeauthoryear{De~Spain}{2014}]{despain2014landscape}
De~Spain, K.
\newblock 2014.
\newblock {\em Landscape of the Now: A Topography of Movement Improvisation}.
\newblock Oxford University Press.

\bibitem[\protect\citeauthoryear{Deterding \bgroup et al.\egroup }{2017}]{deterding2017mixed}
Deterding, S.; Hook, J.; Fiebrink, R.; Gillies, M.; Gow, J.; Akten, M.; Smith, G.; Liapis, A.; and Compton, K.
\newblock 2017.
\newblock Mixed-initiative creative interfaces.
\newblock In {\em Proceedings of the 2017 CHI Conference Extended Abstracts on Human Factors in Computing Systems}, CHI EA '17,  628--635.
\newblock New York, NY, USA: ACM.

\bibitem[\protect\citeauthoryear{Foster}{2002}]{foster2002dances}
Foster, S.
\newblock 2002.
\newblock {\em Dances that Describe Themselves: The Improvised Choreography of Richard Bull}.
\newblock Wesleyan University Press.

\bibitem[\protect\citeauthoryear{Goodfellow \bgroup et al.\egroup }{2014}]{goodfellow2014generative}
Goodfellow, I.; Pouget-Abadie, J.; Mirza, M.; Xu, B.; Warde-Farley, D.; Ozair, S.; Courville, A.; and Bengio, Y.
\newblock 2014.
\newblock Generative adversarial nets.
\newblock In {\em Advances in Neural Information Processing Systems},  2672--2680.

\bibitem[\protect\citeauthoryear{Hoffman and Weinberg}{2011}]{hoffman2011interactive}
Hoffman, G., and Weinberg, G.
\newblock 2011.
\newblock Interactive improvisation with a robotic marimba player.
\newblock {\em Autonomous Robots} 31(2):133--153.

\bibitem[\protect\citeauthoryear{Jennings}{2010}]{jennings2010developing}
Jennings, K.~E.
\newblock 2010.
\newblock Developing creativity: Artificial barriers in artificial intelligence.
\newblock {\em Minds and Machines} 20(4):489--501.

\bibitem[\protect\citeauthoryear{Kingma and Welling}{2014}]{kingma2014auto}
Kingma, D.~P., and Welling, M.
\newblock 2014.
\newblock Auto-encoding variational {B}ayes.
\newblock In {\em Proceedings of the 2nd International Conference on Learning Representations (ICLR)}.

\bibitem[\protect\citeauthoryear{Kumar, Long, and Magerko}{2020}]{kumar2020creativity}
Kumar, M.; Long, D.; and Magerko, B.
\newblock 2020.
\newblock Creativity metrics for a lead-and-follow dynamic in an improvisational dance agent.
\newblock In {\em Proceedings of the International Conference on Computational Creativity}.

\bibitem[\protect\citeauthoryear{Lee \bgroup et al.\egroup }{2019}]{lee2019dancing}
Lee, H.-Y.; Yang, X.; Liu, M.-Y.; Wang, T.-C.; Lu, Y.-D.; Yang, M.-H.; and Kautz, J.
\newblock 2019.
\newblock Dancing to music.
\newblock In {\em Proceedings of the 33rd International Conference on Neural Information Processing Systems},  3586--3596.
\newblock Red Hook, NY, USA: Curran Associates Inc.

\bibitem[\protect\citeauthoryear{Nachmanovitch}{1990}]{nachmanovitch1990free}
Nachmanovitch, S.
\newblock 1990.
\newblock {\em Free Play: Improvisation in Life and Art}.
\newblock Penguin Putnam.

\bibitem[\protect\citeauthoryear{Tang, Jia, and Mao}{2018}]{tang2018dance}
Tang, T.; Jia, J.; and Mao, H.
\newblock 2018.
\newblock Dance with melody: An lstm-autoencoder approach to music-oriented dance synthesis.
\newblock In {\em Proceedings of ACM Multimedia Conference}.

\bibitem[\protect\citeauthoryear{Tsuchida \bgroup et al.\egroup }{2019}]{aist-dance-db}
Tsuchida, S.; Fukayama, S.; Hamasaki, M.; and Goto, M.
\newblock 2019.
\newblock {AIST} dance video database: Multi-genre, multi-dancer, and multi-camera database for dance information processing.
\newblock In {\em Proceedings of the 20th International Society for Music Information Retrieval Conference, {ISMIR} 2019}.

\bibitem[\protect\citeauthoryear{Winship}{2024}]{winship2024small}
Winship, L.
\newblock 2024.
\newblock Small step or a giant leap? what {AI} means for the dance world.
\newblock {\em The Guardian}.
\newblock Accessed: January 11, 2025.

\end{thebibliography}

\end{document}